\documentstyle[aps,preprint]{revtex}
\begin{document}

\title{Nanometer-scale capacitors}
 
\draft
\tightenlines

\author{ J. Stepani\'c  }

\address{Faculty of Mechanical Engineering and Naval Architecture,
University of Zagreb, I. Lu{\v c}i{\' c}a 5,
10000 Zagreb, Croatia}

\author{ G. Bilalbegovi\'c }

\address{Department of Physics, 
University of Rijeka, Omladinska 14, 51000 Rijeka, Croatia}

\date{\today}

\maketitle

\begin{abstract}

Molecular dynamics computer simulations which employ 
the embedded-atom potential show that 
nanowires of gold exist as multishelled 
structures. We simulate double-walled gold nanowires and
calculate the capacitance of a finite nanometer-size cylindrical capacitor.
For the sizes for which multishelled nanowires appear in simulations we
find the capacitances below one attofarad.  
    
\end{abstract}

\pacs{68.65.+g,61.46.+w,73.61.Tm,85.40.Ux}

Finite-size systems of atomic scale dimensions
have attracted great attention. Interest is focused on fundamental
aspects, as well as on applications. The remarkable physical properties of
nanostructures allow one to design nano-scale electrical and mechanical
devices. The continuing miniaturization of engineering devices leads
to a technological revolution. Research on the properties of cylindrical
nanostructures of various materials, ie., nanowires, has been 
especially active field.

Multishelled nanostructures were
produced for carbon clusters and wires \cite{Iijima,Ugarte},
as well as for wires of $WS_2$, $MoS_2$, and $NiCl_2$ \cite{Tenne}.
Gold nanowires are fabricated by Scanning Tunneling Microscopy (STM)
\cite{Ohnishi,Yanson}
and electron-beam litography \cite{Hegger}.
In recent Molecular Dynamics (MD) computer simulation multishelled cylindrical 
nanostructures of gold were obtained \cite{Goranka}. 
Wires with radii around a nanometer and
a length/diameter ratio between $1$ and $3$ were studied. 
It was found that 
multiwalled structures exist for longer of these 
nanowires, i.e., for a length/diameter ratio of $2$ and $3$. 
The formation of shells was also found in a jellium model calculation for 
finite sodium nanowires \cite{Landman}. 
This was confirmed by the conductance measurements \cite{Amsterdam}.
We simulate here 
double-walled gold nanowires and study the capacitance for this
geometry.

It is well known that an application of a many-body potential is necessary
for accurate description of 
metallic bonding by the MD simulation method
\cite{Daw}. We used here (as in Ref. \cite{Goranka}) an embedded-atom
potential for gold which produces a good agreement 
with available experimental results for bulk, surfaces, 
and nanoparticles \cite{Furio}. 
A time step of $7.14\times 10^{-15}$ s was employed in simulation.
The temperature was controlled by rescaling the particle velocities. 
We started from an ideal fcc (111) structure at $T=0$ K, and included in the
cylindrical MD box all particles whose distance from the nanowire axis was
smaller than $1.2$ nm. The length of the sample of 
$689$ atoms was $6$ layers.
We also studied a nanowire of $9$ layers and 1032 particles.
The samples were first relaxed, then annealed and quenched.
To prevent melting and collapse into a drop, 
instead of usual heating to $\sim 1000$ K used in MD simulation of gold 
nanostructures  our finite nanowires were heated only to $600$ K.
Such a procedure, as in Ref. \cite{Goranka},
gives the atoms a possibility to find local minima
and models a constrained dynamical evolution present in fabricated nanowires.
The structures were analyzed after a long MD run at $T=300$ K.
Figure 1 shows the shape of the MD box for a nanowire of $689$ atoms after
$3.55$ ns of simulation. 
A top view of the whole cylindrical MD box for this nanowire 
is shown in  Fig.~\ref{fig2}.
While the presence of a double-walled and internally filled structure is
obvious, the walls are still not
completely homogeneous. Several atoms remain about the walls.
Simulation shows that an almost identical structure exists up to $10^6$
time steps, i.e., $7.1$ ns \cite{MS}. 
We also found a double-walled structure for nanowire of $1032$ atoms. 
These structures suggest an application 
of gold nanowires as cylindrical capacitors. 
Therefore, we calculated the capacitance of a nanometer-scale capacitor
for the geometry shown in the inset of Fig.~\ref{fig3}.

The calculation starts 
from the electrical potential $\phi(r,z)$ in the space between 
two coaxial cylinders of radii $r_1$ and $r_2$ ($r_1 < r_2$), and finite
length $L$ in the $z$ direction \cite{Cirlin} 
\begin{equation}
\label{eq:1}
\phi(r,z)=V\frac{4}{\pi}\sum_{k=1,3,5,...}\frac{\sin\nu_{k}z}{k
}
\frac{I_{0}(\nu_{k}r)K_{0}(\nu_{k}r_{1})-I_{0}(\nu_{k}r_{1})K_{0}(\nu_{k}
r)}
{I_{0}(\nu_{k}r_{2})K_{0}(\nu_{k}r_{1})-I_{0}(\nu_{k}r_{1})K_{0}(\nu_{k}r
_{2})},
\end{equation}
where $\nu_{k}={k\pi}/{L}$, $r_1 \leq r \leq r_2$, and $0 \leq z \leq L$.
The parameter $V$ is the potential difference applied between 
two cylindrical plates, whereas 
$I_{i}(x)$ and $K_{i}(x)$ (here and below)
are the modified Bessel functions of the order 
$i$\cite{gradsteyn}. The capacitance $C=(r_1,r_2,L)$ of the structure is 
obtained from 
\begin{equation}
\label{eq:2}
C=\frac {2\pi r_1 \epsilon_{0}}{V} \int_0^L {\left(\frac {\partial \phi}
{\partial r}\right)_{r=r_1}dz,}
\end{equation}
where $\epsilon_0 = 8,854 \times 10^{-12}$ $F$ $m^{-1}$ is the electric
permittivity of vacuum.
The result is
\begin{equation}
\label{eq:3}
C=\left. 16r_{1}\epsilon_{0}\sum_{k=1,3,5,...}\frac{1}{k}
\frac{
I_{0}(\nu_{k}r_{1})K_{1}(\nu_{k}r_{1})+I_{1}(\nu_{k}r_{1})K_{0}(\nu_{k}r_{1})}
{I_{0}(\nu_{k}r_{1})K_{0}(\nu_{k}r_{2})-
I_{0}(\nu_{k}r_{2})K_{0}(\nu_{k}r_{1})}\right..
\end{equation} 
Using  equation (3) we calculated the capacitance for several sizes typical
for structures obtained in MD simulations.
The results are presented in Fig.~\ref{fig3} and Table 1.

In summary,
molecular dynamics computer simulation based on the well-established potential
shows that gold wires of nanometer dimensions
form double-walled structures. 
Motivated by this result we calculated the capacitance of a finite
nanometer-scale cylindrical capacitor. 
Our calculations of the capacitance is based on the classical electrodynamics.
However, on very small scales the quantum corrections appear. This is the
point we are now pursuing \cite{Josip}.
It is known that typical inhomogeneous electric fields 
used in the polarizability measurements for metallic clusters are on 
the order of $10^5$ $V cm^{-1}$ and field gradients are 
$\sim 10^5$ $Vcm^{-2}$  \cite{deHeer}.  
These fields induce a dipole 
moment and slightly deform clusters. 
Therefore, to strictly preserve 
the shape of a nanometer-scale capacitor weaker and homogeneous fields should 
be applied in an electronic device.  
An STM could be used to manipulate gold
atoms and assemble small nanowires \cite{Ohnishi,Yanson}.
The STM experiments show that gold nanostructures can sustain large
fields of up to $1$ V \cite{Yanson}. 
Wires produced by 
electron-beam litography will approach the
smallest nanometer-scales discussed here.
It is possible to fill  
the space between two cylindrical electrodes  
with various insulators and to obtain 
nanometer-scale capacitors 
with different capacitances. Similar filling of carbon nanotubes with metals
was recently realized \cite{Ajayan,Demoncy}. 
Moreover, fabrication of metallic nanowires using carbon 
multishelled nanotubes as templates could be used to produce
nanometer-scale cylindrical capacitors not only of gold,
but also of other metals.

\acknowledgments

This work has been carried out under 
the HR-MZT project 119206 - ``Dynamical Properties of Surfaces'' and 
the EC Research Action COST P3 - ``Simulation of Physical Phenomena 
in Technological Applications''.
The paper is dedicated to Prof. B. Leonti{\'c} on occasion of his 70th
birthday.

\begin{table}
\caption{Capacitance $C$  of a nanometer-scale capacitor 
with an external
radius $r_2=1.2$ nm, an internal radius $r_1=0.9$ nm, for several
lengths $L$.}
\label{table1} 
\begin{tabular}{l l l l l l l }  
$L (nm)$  & $2.0$ & $2.2$ & $2.4$ & $2.6$  & $2.8$ & $3.0$   \\

$C (aF)$  & $0.336$ & $0.374$ & $0.413$ & $0.452$ & $0.490$ & $0.529$\\
\end{tabular}
\end{table}

\begin{figure}
\caption{
Atomic positions (side view)
for a nanowire with length $2.6$ nm and an external radius
of $1.2$ nm.}
\label{fig1}
\end{figure}

\begin{figure}
\caption{MD trajectory plot of a top view for 
a gold nanowire shown in Fig. 1.
The whole thickness of the wire along its axis is shown
after $3.55$ ns of simulation.
This nanowire consists of the two coaxial near walls and a large filled core.}
\label{fig2}
\end{figure}

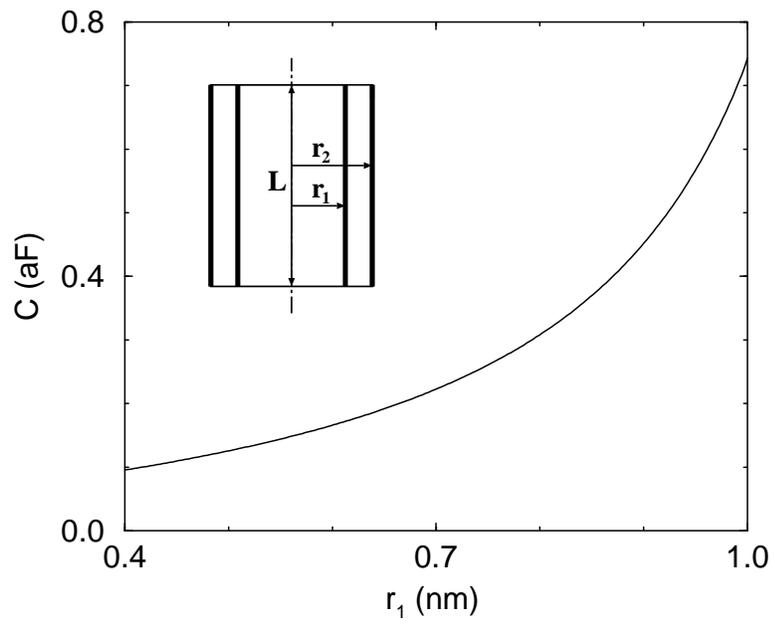
\begin{figure}
\caption{
Capacitance of a capacitor as a function of an internal 
radius $r_1$, for $L=2.6$ nm, and an external radius
$r_2=1.2$ nm. Inset:
the geometry of a nanometer-scale cylindrical capacitor.}
\label{fig3}
\end{figure}

\end{document}